\documentclass[twocolumn,showpacs,amsmath,amssymb,prl,superscriptaddress]{revtex4}
\usepackage{graphicx}
\usepackage{bm}
\usepackage{dcolumn}
\usepackage{color}

\begin{document}
\title{Cooperative effects of Jahn-Teller distortion, magnetism and Hund's coupling \\ in the insulating phase of BaCrO$_3$}

\author{Gianluca Giovannetti}
\affiliation{CNR-IOM-Democritos National Simulation Centre and International School
for Advanced Studies (SISSA), Via Bonomea 265, I-34136, Trieste, Italy}
\author{Markus Aichhorn}
\affiliation{Institute of Theoretical and Computational Physics, TU
  Graz, Petersgasse 16, Graz, Austria}
\author{Massimo Capone}
\affiliation{CNR-IOM-Democritos National Simulation Centre and International School
for Advanced Studies (SISSA), Via Bonomea 265, I-34136, Trieste, Italy}

\begin{abstract}
We employ a combination of density functional theory and dynamical
mean-field theory to investigate the electronic structure of the recently
synthesized insulator BaCrO$_3$. Our calculations show that Hund's
coupling is responsible for strong correlation effects, which are
however not sufficient to turn the system insulating. A finite
Jahn-Teller distortion lifting the orbital degeneracy is necessary to stabilize an insulating state
with orbital ordering and consequent magnetic ordering. 
\end{abstract}

\maketitle

Strongly correlated materials are characterized by narrow valence bands arising from localized atomic orbitals. The reduced itinerancy of the carriers is the gateway that leads to the remarkable variety of phases and regimes observed in d- and f-electron systems.
Usually strong electronic correlations are associated with the proximity of a Mott insulating state when the screened local Coulomb interaction U reaches a critical value U$_c$ comparable with the bandwidth W of the system. For pure Hubbard-like interactions U$_c$ would increase with the orbital degeneracy. However, 
in multi-orbital system, the Hund's exchange does not simply lead to a multiplet splitting, but it qualitatively affects the strength and the nature of the electronic correlations \cite{Georgesreview}. 
If the number of the electrons per correlated atoms is not equal to
the number of orbitals (i.e., away from half-filling), the
Hund's coupling increases $U_c$, but at the same time reduces the
coherence of the system at intermediate interactions. This
  leads to a wide range of interaction parameters where the
carriers are strongly correlated, but at the same time far from a Mott insulating
state. This novel correlated regime is often referred to as a Hund's
metal. 

While fingerprints of Hund's metal physics have been reported to
explain the bad metallic behavior of iron-based
superconductors\cite{hund_iron} and ruthenates\cite{hund_ru}, the
interplay of Hund's correlations with electron-lattice coupling and
the relationship with magnetic ordering have not been studied
  in detail. These effects are indeed strongly intertwined. For example,
lattice distortions can lead to crystal-field splitting, which lifts
the orbital degeneracy and may lead to orbital
  selective physics \cite{demediciPRL,demediciIRON}. In addition,
  as we will show below,
crystal-field splitting can compete with Hund's coupling and Hund's
metallicity, since it can
induce orbital polarization, whereas Hund's coupling tends to make orbital
occupancies equal.
Furthermore, strong correlation physics is usually
associated to the formation of local magnetic moments, which typically
order in magnetic patterns which depend on both the interactions and
the lattice structure, eventually leading to insulating behavior. 

The perovskite compounds ACrO$_3$ (A being Ca, Sr, Ba) are a good
playground to study this interplay. In those systems Cr has a nominal
valence of 4+, with 2 electrons in the three-fold degenerate t$_{2g}$
orbitals. Interestingly, such an electronic configuration is common to
both metallic systems like (Ca,Sr)RuO$_3$ and insulating vanadates
with formula RVO$_3$ (R being rare-earth), in which different orbital
and spin orderings establish as a function of temperature \cite{RVO3}.  

Recently resistance measurements revealed BaCrO$_3$ to be insulating
over a large range of temperature with a low-temperature gap of
0.38\,eV \cite{BaCrO3exp}, while there are conflicting evidences of
metallic and insulator behaviors in CaCrO$_3$ and SrCrO$_3$
\cite{Zhou,Chamberland,Williams,Komarek} and the general question of
whether these compounds are metals or insulators is still under
debate.  

The compound with the smallest A-site ion in the series, CaCrO$_3$,
shows an orthorhombically distorted GdFeO$_3$ structure in which the three
Cr-O distances are comparable to those found in YVO$_3$ at room
temperature \cite{Zhou} and is characterized by a C-type
antiferromagnetic (AFM) magnetic structure \cite{Komarek}. The same
C-type magnetism has been reported in SrCrO$_3$ in which
however the octahedral-like GdFeO$_3$ distortions are suppressed \cite{Komarek}.  
Density-functional theory (DFT) calculations, including interaction
effects within a DFT+U scheme, found CaCrO$_3$
\cite{Streltsov} and SrCrO$_3$ \cite{Lee} to be either weakly metallic or with
a small gap, exhibiting C-type
antiferromagnetism with orbital ordering.

To properly account for the delicate balance between inherent
correlation effects and the tendency towards ordering, we use a
combined approach based on DFT \cite{DFT} and dynamical mean-field
theory (DMFT) \cite{DMFT} which accurately treats the electronic
correlations. Our study accounts for the insulating state of
BaCrO$_3$, highlighting the role of Coulomb interactions $U$ and $J$ and
their interplay with the lattice distortions. 

We start our investigation with  DFT calculations within the
local-density approximation (LDA)
for the tetragonal unit cell of BaCrO$_3$ using Quantum Espresso
\cite{QE} and Wien2K \cite{Wien2k} packages. We build Wannier
orbitals \cite{wannier90,TRIQS} for the t$_{2g}$ (xy,xz,yz) manifold of Cr,
which define the basis where we consider  the 
interaction effects. The Coulomb interaction within the t$_{2g}$
manifold is taken in the Kanamori form 
\begin{align*}
H&_{int} =   U\sum_{i,m}n_{im{\sigma}} n_{im{\sigma}'} +U'\sum_{i,m,m'}n_{im{\sigma}} n_{im'{\sigma}'} +\\
&+U^{''}\sum_{i,m,m'}n_{im{\sigma}} n_{im'{\sigma}} \\
& -J_h \sum_{i,m,m'} [ d^{+}_{im{\uparrow}}d^{+}_{im'{\downarrow}}d_{im{\downarrow}}d_{im'{\uparrow}} + d^{+}_{im{\uparrow}}d^{+}_{im{\downarrow}}d_{im'{\uparrow}}d_{im'{\downarrow}}   ],\\
\end{align*}
where $d_{i,m{\sigma}}$ is the annihilation operator of an
electron of spin $\sigma$ at site $i$ in orbital $m$, and $n_{im{\sigma}}=
d^{+}_{im{\sigma}}d_{im{\sigma}}$ is the density operator. Intra- and
inter-orbital repulsions are given by $U$, $U'=U-2J_h$, and
$U^{''}=U-3J_h$, and $J_h$ is the Hund's coupling.

The values of $U$ and $J_h$ are determined on the basis of constrained
random-phase approximation (cRPA) calculations that have been
performed for the similar
material SrCrO$_3$ \cite{cRPASrMO3}, which gave $U=2.7$\,eV and
$J_h=0.42$\,eV. However the lattice parameters of SrCrO$_3$ are smaller
with respect to those of BaCrO$_3$ by 0.1\,$\AA$ so that we expect Cr d
orbitals will be more localized in our compound. For this reason we
use a slightly larger $U$ keeping the same ratio $J{_h}/U$ of SrCrO$_3$,
namely $U=3.25$\,eV and $J_h=0.505$\,eV. 
DMFT requires the numerical solution of a quantum impurity model
including the local interactions. We use two implementations
  of LDA+DMFT, one using Exact Diagonalization (ED)
  \cite{Caffarel,Capone}, the other one using continuous
time quantum Monte Carlo (CTQMC) \cite{TRIQS,Gull2011,legendre} to solve for the local Green's function of the impurity model implementing the conservation laws introduced in Ref.\cite{PS}. 
In order to study phases with magnetic and/or orbital ordering which
spontaneously break the translational symmetry of the lattice we
introduce a quantum impurity solver for each inequivalent Chromium site
(or sublattice). This is equivalent to assume a local spin- and
orbital-dependent self-energy with different values on each
inequivalent site. 

\begin{figure}
\includegraphics[width=.75\columnwidth,angle=-0]{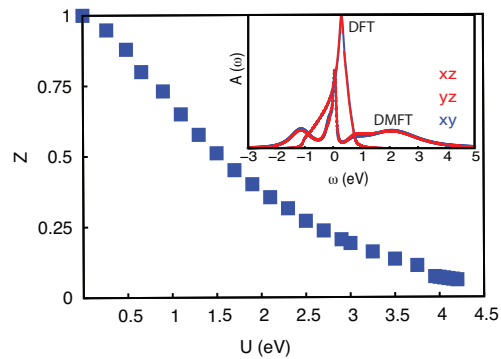}
\caption{(Color online) Quasiparticle weight Z as function of U fixing the ratio $J{_h}/U$ fixed to the value 0.155 calculated with ED solver. Inset: Resolved orbital density of state of Wannier orbitals (DFT) and DMFT calculations.}
\label{fig1}
\end{figure}

In the tetragonal unit cell of BaCrO$_3$ \cite{BaCrO3exp}
(a$=$b$=$4.09 $\AA$, c$=$4.07 $\AA$) the small deviation from a
pseudo-cubic perovskite structure lifts the orbital degeneracy of the
t$_{2g}$ orbitals, shifting the xy orbital to lower energy with
respect to the degenerate (xz,yz) orbitals. The crystal-field
splitting is 0.02 eV at LDA level. 
Non-magnetic LDA calculations describe  BaCrO$_3$ as a metal, as
can be clearly seen from the orbitally-resolved density of
  states shown in the inset of Fig. \ref{fig1}. All the three bands
arising from the t$_{2g}$ orbitals  cross the Fermi level and they share a
similar bandwidth of around 1.7 eV, similarly to the other compound of
the same family SrCrO3 \cite{cRPASrMO3}. In particular the filling of
the chromium bands confirms that the bandstructure is associated to
Cr$^{4+}$ ions and to a total filling of two electrons populating the
three t$_{2g}$ orbitals.  

Since the crystal field splitting is small compared to the
  bandwidth, the orbital occupancies in our calculations remain
nearly degenerate being close to 0.66 electrons per orbital. 

When we include the interaction effects in LDA+DMFT we expect a
reduction of the effective bandwidth for the coherent motion of the
carriers. The quasiparticle weight $Z$ is a quantitative measure of this
effect. A small value of $Z$ is a fingerprint of a strongly correlated
state, and a vanishing $Z$ marks a Mott-Hubbard metal-insulator
transition. In Fig. \ref{fig1}, we display our results for $Z$-which is
independent on the orbital index-in the paramagnetic state as a
function of interaction $U$, keeping the ratio of $J_h/U$
fixed at the cRPA value 0.155.  

The evolution of the quasiparticle weight $Z$ displays the typical
Hund's metal behavior, with a clear change of curvature around
$U\approx 1.5$\,eV, separating a
fast drop in the weak-coupling range from a flatter region for
intermediate coupling in which $Z$ is small but finite\cite{DeMediciJanus}. 
The Mott phase is indeed reached only at values of $U$ close to
5\,eV, significantly larger than the cRPA estimate.

\begin{figure}
\includegraphics[width=.95\columnwidth,angle=0]{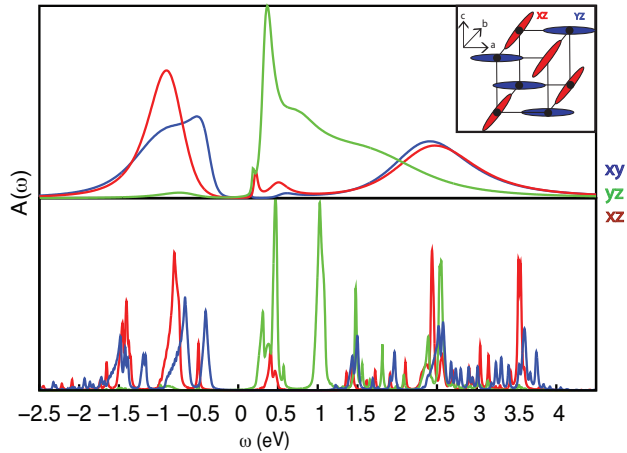}
\caption{(Color online) Resolved orbital spectral density for
  $U=3.25$\,eV, $J_h=0.505$\,eV in G-type OO for $\delta=0.17 \AA$. Lower panel shows
  results using the ED solver, the upper panel results from CTQMC.
  Inset: Orbital arrangement (G-type OO) in the plane of tetragonal unit cell
  with the two inequivalent Cr-O distances. }
\label{fig2}
\end{figure}

Our calculations show that pure correlations effect, including the
Hund's exchange, do not result in an insulating behavior that
is experimentally observed in BaCrO$_3$ \cite{BaCrO3exp}, and that other physical effects
have to be taken into account. The most straightforward mechanism to turn the Hund's metal into an
insulator invokes magnetic ordering, but this simple scenario seems to
be ruled out by local-spin-density approximation (LSDA) calculations
including an Hartree-Fock treatment of interactions (LSDA+U) which indeed stabilize a C-type
magnetic structure that however remains metallic \cite{BaCrO3exp}.  

As mentioned above, another crucial effect to tune electronic correlations
is the crystal-field splitting, which can be a consequence of a
cooperative Jahn-Teller (JT) effect \cite{JahnTeller} in which the lattice
structure is distorted and the orbital degeneracy is lifted. It
should be noted that a JT-driven splitting competes with the
Hund's coupling concerning the orbital occupancies. 
Whereas the former tends to
populate most the lowest-lying orbitals, the latter tends
to distribute electrons equally among the orbitals to reduce the Coulomb
repulsion by the intra-atomic exchange.

In electronically similar vanadates LaVO$_3$ and YVO$_3$ the
JT distortion of the octahedra has indeed a dramatic
influence on their electronic properties \cite{Blake,Fang}. The role
of JT distortions in isoelectronic Chromium compounds has not
been discussed to our knowledge, but LSDA+U calculations for CaCrO$_3$
predict orbital ordering in which one electron occupies the
orbital with mostly xy character and the second one occupies the two
combinations $xz+yz$ and $xz-yz$ in a staggered pattern
\cite{Streltsov}, a configuration that would be further favored by 
JT distortions. 

\begin{figure}
\includegraphics[width=.95\columnwidth,angle=0]{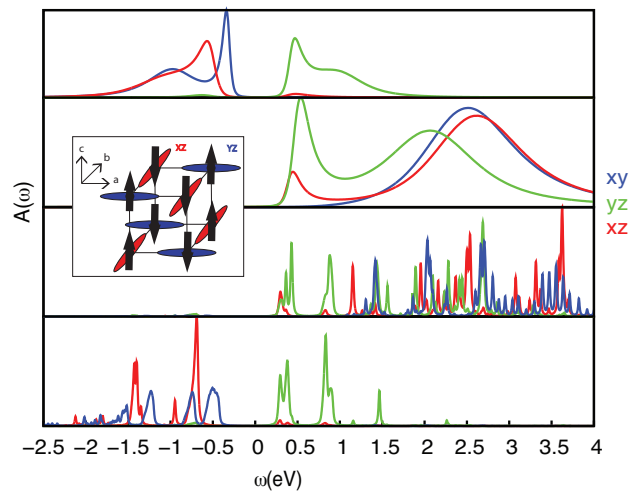}
\caption{(Color online) Orbitally-resolved spectral density
  for $U=3.25$\,eV, $J_h=0.505$\,eV in G-type orbital ordering and C-type spin ordering
  for $\delta=0.17 \AA$. 
  Lower two panels: Results using the ED solver for Majority/Minority spin channels. 
  Upper two panels: Results from CTQMC for Majority/Minority spin channels.
  Inset: Spin (C-type AFM) and Orbital (G-type OO) arrangement in the plane 
  of tetragonal unit cell with the two inequivalent Cr-O distances being.} 
\label{fig3}
\end{figure}

On the basis of the evidence for similar compounds, we can expect the
JT effect in BaCrO$_3$ to give rise to two different possible
orbitally-ordered configurations. First, we can have orbital
  ordering (OO) of the xz and yz orbitals in G-type fashion, where xz
  and yz orbitals alternate in all three spatial dimensions (see Inset
Fig. \ref{fig2}). The second possibility is C-type OO, where xz and yz order in AF
  fashion in the ab plane, but ferromagnetically along the c-axis.
We perform different LDA+DMFT calculations at different values of the
short/long Cr-O distances ($\delta$) with G-type and C-type OO as allowed by
symmetry considerations. The results weakly depend on the value of
$\delta$ and we present calculations for $\delta$=0.17 $\AA$, which is
a reasonable value for BaCrO$_3$ (See below).

The distortion of the oxygen sites splits the position of the
energy levels of the (xy,xz,yz) orbitals: at a particular Cr site the
xy lies between xz and yz orbitals which are higher and lower in energy by an
amount of 0.05\,eV with the two inequivalent Cr-O distances in the plane
of tetragonal unit cell being the above mentioned $\delta=0.17 \AA$.  
The non-interacting orbital density of states shows three bands with
nearly same bandwidth and shape of 1.8 eV with orbital occupancies
being 0.79, 0.66 and 0.55 from the lower to higher energy level
(xz,xy,yz), respectively. 
According to the symmetry of the crystal the occupancies of the xz and yz
orbitals alternate in the a-b plane. The above results are
essentially independent on the choice of C-type or G-type orbital
ordering.

In this lattice configuration DMFT calculations enhance the
occupancies of xz and xy orbitals with respect to yz upon increasing the
electronic correlations as if the crystal-field splitting is
effectively increased. Interestingly, the effect is realized via a
strongly $\omega$-dependent real part of the self-energy.

At $U=3.25$\,eV and $J_h=0.505$\,eV in the paramagnetic state the orbital
occupancies of t$_{2g}$ orbitals in the insulating solutions are 0.96,
1.0 and 0.04 for xz,xy,yz respectively, showing that correlations enhance
the orbital correlations and result in almost perfect orbital ordering.
The resolved orbital spectral density with the chosen values of $U$
and $J_h$ in Fig. \ref{fig2} show that with the inclusion of a
JT distortion the system turns insulating with a gap in good
agreement with experimental values \cite{BaCrO3exp}. 

The relative orientation of orbitals on neighboring cations then determines the superexchange interactions between the unpaired d-electrons, which in turn determine the magnetic ordering patterns. 
The presence of a JT distortion is expected to give rise to a strong interplay between orbitals and spins \cite{KugelKhomskii}.For perovskites with partially occupied e$_g$ orbitals, such
as cuprates \cite{KCuF3} and manganites \cite{LaMnO3} or other Cr based compounds \cite{KCrF3},
the JT energy greatly exceeds the superexchange coupling between
unpaired electrons, so that orbital and magnetic ordering temperatures
are well separated. In the case of BaCrO$_3$, where only t$_{2g}$
orbitals contribute to low-energy physics, we expect a weaker
JT coupling and orbital and magnetic transitions could occur in the same
temperature range. The orbital and spin orderings in BaCrO$_3$ might develop the same temperature-induced magnetization reversal \cite{Palstra} and different ordering temperatures \cite{RVO3} as in YVO$_3$ single crystal.

The calculated orbital occupancies establish that the lattice
distortion triggers the magnetic ordering with a C-type
(each spin is antiparallel to all others in the ab plane but parallel
along the c axis) or G-type (every spin is antiparallel to all its
neighbors) magnetic structure stabilized depending if a G-type/C-type
OO is considered as expected from the Goodenough-Kanamori rule for
superexchange \cite{GoodenoughKanamori}.  
In the magnetically ordered states the (xz,xy,yz) orbitals are filled by
0.95(0.0), 0.98(0.02), 0.03(0.02) electrons in the majority(minority)
channels respectively. Once again, these populations do not depend on
the precise nature of the ordered state.
Magnetic ordering further increases the value of the gap by
0.1\,eV as can be seen form the resolved orbital spectral density in 
Fig. \ref{fig3} for the case of C-type magnetism and G-type OO. 

We mention here that within LDA+DMFT we discard the possibility of a
octahedral tilting 
with a GdFeO$_3$ type of distortion as in the case of CaCrO$_3$ crystal structure being the ionic radius of Ba larger as for Sr.
Moreover a combination JT distortions and tilting of
the oxygen octahedra allows by symmetry for many different space
groups going beyond the scope of this work.
There are only few attempts within LDA+DMFT schemes for fully microscopic
investigations of the structural properties of strongly correlated
electron materials such as lattice instabilities \cite{Leonov}. To
further check our theoretical prediction we perform DFT+U calculations
using the Wien2k package and an effective interaction
$U_{eff}=U-J=4$\,eV. Note that interactions are here applied to atomic
orbitals, and therefore the value for the Coulomb interaction
is larger. 

\begin{figure}
\hspace{-2.0cm}\includegraphics[width=.95\columnwidth,angle=0]{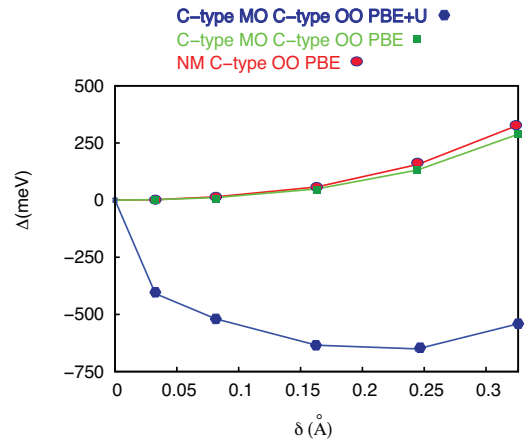}
\caption{(Color online) Total energy gain calculated within DFT in the nonmagnetic (NM) and C-type magnetic structure and within DFT+U in the C-type magnetic structure as function of the Jahn-Teller distortion $\delta$ (difference between long/short Cr-O bonds).}
\label{fig4}
\end{figure}

In Fig. \ref{fig4} we show the total energy gain as
function of the JT distortion for the non-magnetic, C-type magnetic
structure within DFT and C-type magnetic structure within DFT+U, which
confirm that the presence of a JT instability and the role of the
Coulomb interaction in the insulating state of 
BaCrO$_3$. However, LDA+DMFT paramagnetic calculations provide an
insulating solution for every finite JT distortion considered, while
DFT+U requires the inclusion of magnetism to open the insulating gap.

In conclusion, we use an LDA+DMFT approach to study the electronic
properties of the recently synthesized compound BaCrO$_3$, which is a
non trivial insulator. We show that BaCrO$_3$, in its tetragonal crystal
structure, would be a strongly correlated metal in which the degree of
electronic correlations is tuned by the Hund's coupling. We predict
that the insulating character arises from the combination of the above
correlation effects and a Jahn-Teller splitting of the t$_{2g}$ orbitals, which leads to
orbital ordering and to a gap in good agreement with the experimental
value. Magnetic ordering compatible with the Goodenough-Kanamori rule
establishes in the JT distorted insulating state. 
Experimental investigations of the crystal structure of BaCrO$_3$
would be a direct test of our scenario. 

We wholeheartedly thank S. Streitsov and L. de' Medici for precious discussions.
GG and MC acknowledge financial support by European Research Council
under FP7/ERC Starting Independent Research Grant ``SUPERBAD" (Grant
Agreement n. 240524). MA acknowledges support from the SFB VICOM
sub-project F04103. Calculations have been performed on the TU Graz
{\em dcluster} and at CINECA (HPC project lsB06\_SUPMOT).

\end{document}